\pdfoutput=1
\documentclass[aps,prl,twocolumn,10pt]{revtex4-1}
\usepackage[utf8]{inputenc}
\usepackage{fourier}
\usepackage{amsmath,graphicx,microtype,bm}
\usepackage[cal=boondoxo,frak=euler]{mathalfa}
\usepackage[colorlinks=true,citecolor=blue,urlcolor=blue]{hyperref}

\begin{document}

\title{Dynamics of one-dimensional electrons with broken spin-charge separation}

\author{Yasha Gindikin and Vladimir A.\ Sablikov}

\affiliation{Kotel'nikov Institute of Radio Engineering and Electronics,
Russian Academy of Sciences, Fryazino, Moscow District, 141190, Russia}

\begin{abstract}
Spin-charge separation is known to be broken in many physically interesting one-dimensional (1D) and quasi-1D systems with spin-orbit interaction because of which spin and charge degrees of freedom are mixed in collective excitations. Mixed spin-charge modes carry an electric charge and therefore can be investigated by electrical means. We explore this possibility by studying the dynamic conductance of a 1D electron system with image-potential-induced spin-orbit interaction. The real part of the admittance reveals an oscillatory behavior versus frequency that reflects the collective excitation resonances for both modes at their respective transit frequencies. By analyzing the frequency dependence of the conductance the mode velocities can be found and their spin-charge structure can be determined quantitatively.
\end{abstract}

\maketitle

\textit{Introduction}.---Spin-orbit interaction (SOI) causes a range of non-trivial effects in low-dimensional electron systems, especially if combined with electron-electron (e-e) interaction~\cite{manchon2015new}. Below we investigate one yet little studied aspect of SOI in one-dimensional (1D) and quasi-1D systems. Owing to the e-e interaction 1D electrons form a strongly correlated state known as the Tomonaga-Luttinger liquid, the hallmark of which is a spin-charge separation (SCS)~\cite{0034-4885-58-9-002}. The SCS was studied in detail in systems without SOI\@. In the presence of SOI the SCS is still respected in strictly 1D systems. However, the SCS is violated in realistic quasi-1D structures with  transverse quantization sub-bands since the spin is no longer a good quantum number there, resulting in new collective excitations modes, in which spin and charge degrees of freedom are mixed~\cite{PhysRevB.62.16900}.

Even more interesting effects accompanied by the SCS violation appear in 1D electron systems with the spin-dependent e-e interaction. This happens when a 1D electron system is placed close to a metallic gate. The electric field of the image charges that electrons induce on the gate gives rise to the image-potential-induced spin-orbit interaction (iSOI), which produces a spin-dependent contribution to the e-e interaction Hamiltonian. The iSOI not only breaks the SCS, but also leads the system to the instability for sufficiently strong interaction~\cite{PhysRevB.95.045138}.

The SCS can also be violated in 1D edge states of two-dimensional topological insulators. These states are known to have a helical structure with the spin locked to the electron momentum. In the simplest commonly studied case of the \(S_z\) symmetry when the spin orientation depends only on the momentum direction but not on its magnitude, the SCS is respected~\cite{PhysRevB.92.195414}. However, the \(S_z\) symmetry is not an inherent property of the topological insulator. Generally, the \(S_z\) symmetry is violated by the SOI~\cite{Konig766,doi:10.1143/JPSJ.77.031007,PhysRevB.77.125319,1367-2630-12-6-065012}. The single-particle states are then modeled as the Kramers pair of 1D states with the spin orientation depending on the momentum magnitude rather then on its direction alone~\cite{PhysRevLett.108.156402,PhysRevB.91.245112}. The packet composed of such states does not possess a definite spin and, which is particularly interesting, the e-e interaction becomes effectively spin-dependent as can be seen from the unusual form of the time-reversal invariant interaction Hamiltonian (see Eq.~(6) in Ref.~\cite{PhysRevLett.108.156402}).

A similar situation occurs in quantum Hall systems at filling factor \(2\) where the frequency dependent a.c.\ conductance measurements reveal mixing of the charge and neutral (spin) collective modes in 1D chiral edge channels~\cite{Bocquillon2013}.

Therefore a sufficiently general problem appears of how to identify the SCS violation in many 1D- and quasi-1D systems, which attract presently a great interest, and to study the mixed-spin-charge collective excitations. The goal of the present paper is to show that this problem can be solved by means of pure electrical measurements of dynamic conductance of a finite 1D system. This becomes possible because both collective modes excited in a system with broken SCS convey the electric charge and thus contribute to the electric response of the system. This is in contrast to the case of the conserved SCS, where the spin modes are observed in the state-of-the-art magneto-tunneling experiments on the array of quantum wires~\cite{Jompol597} or can be detected in no less complicated but not yet realized measurements of the time-resolved dynamics of the spin-polarized density~\cite{0295-5075-86-5-57006}. 

With this goal in mind, we have investigated the admittance of a 1D quantum wire coupled to leads to show that its frequency dependence reveals the characteristic features of the collective excitation spectra that allow one to extract the velocities of both modes and determine their spin-charge structure.

\textit{The model}.---The particular calculations are performed for a finite 1D quantum wire with the iSOI due to the electron image charges induced on a metallic gate. For simplicity, the gate is supposed to be unbiased, so there is no SOI independent of the electron density. We assume that the iSOI magnitude is not too large so that the system is stable.
	
\textit{Hamiltonian}.---The Hamiltonian of 1D electrons in the wire is
\begin{equation}
	H = H_\mathrm{kin} + H_\mathrm{e-e} + H_\mathrm{iSOI}\,.
\label{fullham}
\end{equation}

Here the kinetic energy is given by
\begin{equation}
\label{kin}
	H_\mathrm{kin} = \sum_{s} \int \psi^+_{s}(x)\frac{p_x^2}{2m}\, \psi_{s}(x)\, dx\,,
\end{equation}
with \(\psi_{s}(x)\) being the electron field operator,  \(s = \pm 1\) the spin index, and \(p_{x}\) the momentum operator. The \(x\) axis is directed along the wire, which extends from \(-L/2\) to \(L/2\), and \(y\) axis is directed normally towards the gate.

The e-e interaction energy reads as
\begin{equation}
\label{Hee}
	\begin{split}
		H_\mathrm{e-e} = &\frac{1}{2} \sum_{s_1 s_2} \int 
		 \psi^+_{s_1}(x_1) \psi^+_{s_2}(x_2) U(x_1,x_2) \\ 
		&\! \times \psi_{s_2}(x_2) \psi_{s_1}(x_1)\,dx_1 dx_2\,.
	\end{split}
\end{equation}
The e-e interaction potential \(U(x_1,x_2)\), screened by the image charges, is assumed short-ranged, \(U(x_1,x_2) = U \delta(x_1 - x_2)\).

The iSOI Hamiltonian equals~\cite{PhysRevB.95.045138}
\begin{equation}
\label{isoi}
	\begin{split}
		H_\mathrm{iSOI} = &\frac{\alpha}{2\hbar} \sum_{s_1s_2}  \int \psi^+_{s_1}(x_1) \psi^+_{s_2}(x_2) \left[ E(x_1,x_2)\mathcal{S}_{12} \right. \\
		&{}+ \left. \mathcal{S}_{12} E(x_1,x_2) \right] \psi_{s_2}(x_2) \psi_{s_1}(x_1)\,dx_1 dx_2\,,
	\end{split}
\end{equation}
with \(\mathcal{S}_{12} = (p_{x_1} s_1 + p_{x_2} s_2)/2\) and the SOI constant \(\alpha \). Here \(E(x_1,x_2)\) is the \(y\) component of the electric field acting on an electron at point \(x_2\) from the electron image charge at point \(x_1\). If the distance \(d\) between the gate and the wire is small (\( k_F d \ll 1\), \(k_F\) being the Fermi wave vector), the electric field can be approximated as \(E(x_1,x_2) = E \delta(x_1 - x_2)\). Eq.~\eqref{isoi} together with Eq.~\eqref{Hee} represents a spin-dependent pair interaction Hamiltonian.

The leads are assumed to be 1D non-interacting equipotential conductors~\cite{PhysRevB.52.R5539,PhysRevB.52.R17040,PhysRevB.58.13847}. The potential difference \(V \exp(-i \Omega t)\) applied to the leads drops symmetrically across the contact regions between the gated quantum wire and the leads, so that the external potential is 
\begin{equation}
\label{extpot}
	\varphi_\mathrm{ext}(x,t) = \frac{V}{2} \left[\theta\left(-x - \frac{L}{2}\right) - \theta\left(x - \frac{L}{2}\right)\right]e^{-i \Omega t}\,. 
\end{equation}
The Hamiltonian of electron system in the leads includes \(H_\mathrm{kin}\) and
\begin{equation}
	H_\mathrm{ext} = -e \sum_{s} \int \psi^+_{s}(x)\varphi_\mathrm{ext}(x,t) \psi_{s}(x)\, dx\,.
\end{equation}

The Hamiltonian can be bosonized in a standard way~\cite{haldane1981luttinger}. The field operator of the chiral fermions is presented in the form
\begin{equation}
	\Psi_{rs}(x) = \frac{F_{rs}}{\sqrt{2 \pi \epsilon}} e^{irk_F x} e^{i \phi_{rs}(x)}\,,
\end{equation}
where \(r = \pm 1\) specifies the branch of the linear dispersion, \(\epsilon \) is an ultraviolet cut-off, \(F_{rs}\) is a ladder operator, and \(\phi_{rs}\) is a bosonic phase given by
\begin{equation}
	\phi_{rs} = \frac{2 \pi i r}{L} \sum_{q \ne 0} \frac{e^{-i q x}}{q} \rho_{rs}(q) \,,
\end{equation}
with \(\rho_{rs}(q)\) being the normal ordered (::) fermionic density.

Then
\begin{align}
	H_\mathrm{kin} &= \frac{\hbar v_F}{4 \pi} \sum_{r s} \int : {(\partial_x \phi_{rs})}^2\colon dx\,,\\
	H_\mathrm{e-e} &= \frac{U}{8 \pi^2} \sum \limits_{\substack{r_1 r_2\\s_1 s_2}} r_1 r_2 \int  : \partial_x \phi_{r_1 s_1} \partial_x \phi_{r_2 s_2}\colon dx\,,\\
	H_\mathrm{iSOI} &= \frac{\alpha k_F E}{4 \pi^2} \sum \limits_{\substack{r_1 r_2\\s_1 s_2}} r_1 s_2 \int  : \partial_x \phi_{r_1 s_1} \partial_x \phi_{r_2 s_2}\colon dx\,,\\
	H_\mathrm{ext} &= - \frac{e}{2 \pi} \sum_{rs} r \int \varphi_\mathrm{ext}(x,t) \, \partial_x \phi_{rs}\, dx\,.
\end{align}

\textit{Equation of motion}.---The dynamics of electrons in a quantum wire driven by an external ac-potential is described by the equation of motion for the bosonic phase \(\bm{\phi} = {(\phi_{++},\phi_{+-},\phi_{-+},\phi_{--})}^\intercal \), which reads as
\begin{equation}
\label{eqm}
	\mathbf{A} \partial_x \bm{\phi} =  i\omega \bm{\phi} \,,
\end{equation}
with 
\begin{equation}
	\mathbf{A} =\left(\begin{smallmatrix}
					1 + \mathcal{U} + \mathcal{E} && \mathcal{U} && -\mathcal{U} && -\mathcal{U} -\mathcal{E}\\
					\mathcal{U} && 1 + \mathcal{U} -\mathcal{E} && -\mathcal{U} + \mathcal{E}&& -\mathcal{U}\\
					\mathcal{U} && \mathcal{U} -\mathcal{E} && -1 -\mathcal{U} + \mathcal{E}&& -\mathcal{U}\\
					\mathcal{U} + \mathcal{E} && \mathcal{U} && -\mathcal{U} && -1 -\mathcal{U} -\mathcal{E}
				\end{smallmatrix}\right)\,.
\end{equation}
Dimensionless variables are as follows. Introduce the electron transit time \(\tau = L/v_F \). Then \(\omega = \Omega \tau \), \(\mathcal{U} = U/h v_F \), \(\mathcal{E} = \alpha k_F E/ \pi \hbar v_F \), and \(x \) is normalized over \(L \). 

In the leads the equation of motion takes the form
\begin{equation}
	 r \partial_x \phi_{rs} =  i \omega \phi_{rs} - \frac{\mathcal{V}}{2} \mathrm{sign}(x)\,, 
\end{equation}
with \(\mathcal{V} = e V \tau/ \hbar \).
The solutions are chosen so as to describe the collective excitations propagating away from the quantum  wire region, where they are generated. They are, respectively,
\begin{equation}
\phi_{rs} = i \frac{\mathcal{V}}{2 \omega}+
		\begin{cases}
		    \xi_s\, e^{-i \omega x}, & \text{for } r = -1 \\
		    0, & \text{for } r = +1
		\end{cases}
\end{equation}
in the left lead (\(x < -1/2\)) and
\begin{equation}
\phi_{rs} = -i \frac{\mathcal{V}}{2 \omega}+
		\begin{cases}
		    0, & \text{for } r = -1 \\
		    \zeta_s\, e^{i \omega x}, & \text{for } r = +1
		\end{cases}
\end{equation}
in the right lead (\(x > 1/2\)).

Using continuity conditions for \(\bm{\phi}\) at \(x = \pm 1/2\) we arrive at the boundary conditions for Eq.~\eqref{eqm}:
\begin{equation}
\label{bc}
	\begin{split}
		\phi_{++} \big|_{x = - \frac12} &= \phi_{+-} \big|_{x = - \frac12} = i \frac{\mathcal{V}}{2 \omega}\,,\\
		\phi_{-+} \big|_{x = \frac12} &= \phi_{--} \big|_{x = \frac12} = -i \frac{\mathcal{V}}{2 \omega}\,.
	\end{split}
\end{equation}

The solution of Eq.~\eqref{eqm} is
\begin{equation}
\label{expansion}
	\bm{\phi} = \sum_{i = 1}^4 C_i \bm{h}_i e^{i \frac{\omega}{\lambda_i} x} \,,
\end{equation}
with constants \(C_i\) determined in accordance with Eq.~\eqref{bc}. The eigenvalues \(\lambda_i\) of the matrix \(\mathbf{A}\), corresponding to the eigenvectors \(\bm{h}_i\), are equal to \(\pm \lambda \) and \(\pm \Lambda \), where
\begin{align}
	\lambda = \sqrt{1 + 2\mathcal{U} - 2 \sqrt{\mathcal{U}^2 + \mathcal{E}^2}}\,,\\
	\Lambda = \sqrt{1 + 2\mathcal{U} + 2 \sqrt{\mathcal{U}^2 + \mathcal{E}^2}}\,.
\end{align}
They are just the dimensionless velocities of the collective modes in the wire, as can be seen from Eq.~\eqref{expansion}. The dispersion equations for both branches of collective excitations are \(\omega_{\lambda} = \lambda q\) and \(\omega_{\Lambda} = \Lambda q\).

\textit{Structure of excitations}.---If the iSOI is absent (\(\mathcal{E} = 0\)), the excitations exist separately in the charge and spin sectors. In the case of repulsive e-e interaction, the plasmon velocity \(\Lambda > 1\) is enhanced by the interaction, whereas the spinon velocity \(\lambda = 1\) is not renormalized.

The iSOI breaks the spin-charge separation between the modes. As a result, the modes acquire a complex spin-charge structure that evolves with the change in the iSOI magnitude. Generally speaking, for \(\mathcal{E} \ne 0\) both modes convey charge and spin intertwined, and both contribute to the charge transport. 

The spin-charge structure of the excitations is quantitatively described by a spin-charge separation parameter \(\xi \), defined for each branch as~\cite{PhysRevB.95.045138}
\begin{equation}
	\xi_{\lambda (\Lambda)} = \frac{n^{+}_{q \omega} + n^{-}_{q \omega}}{n^{+}_{q \omega} - n^{-}_{q \omega}}\Bigg|_{\omega = \omega_{\lambda (\Lambda)}}\,.
\end{equation}
Here \(n^{s}_{q\omega}\) is the Fourier component of electron density with spin index \(s\), wave-vector \(q\), and frequency \(\omega \), composing the corresponding collective excitation. A purely spin excitation corresponds to \(\xi = 0\), whereas a purely charge excitation is described by \(\xi \to \infty \). It is important that \(\xi \) is directly determined by the excitation velocities. Thus,
\begin{equation}
	\xi_{\lambda} = \frac{1}{\lambda}\sqrt{\frac{1 - \lambda^2}{\Lambda^2 - 1}}
\end{equation}
and
\begin{equation}
	\xi_{\Lambda} = \frac{1}{\Lambda}\sqrt{\frac{\Lambda^2 - 1}{1 - \lambda^2}}\,.
\end{equation}

\textit{Admittance}.---The \(x\)-dependent electron current in the wire obtained from the continuity equation for the electron density equals
\begin{equation}
	j_{\omega}(x) = -\frac{i e \omega}{2 \pi} \sum_{r s} r \phi_{rs}(x)\,.
\end{equation}
According to the Shockley theorem~\cite{doi:10.1063/1.1710367,PhysRevB.58.13847}, the observable current defined as a charge flow through the leads is given by
\begin{equation}
	J = \frac{1}{V} \int_{-L/2}^{L/2} j(x) E_\mathrm{ext}(x)\, dx\,,
\end{equation}
with \(E_\mathrm{ext}(x) = - \nabla \varphi_\mathrm{ext}\) being the external field along the electron trajectory. The trivial capacitive current between the leads is disregarded.

The admittance \(G_{\omega} = J_{\omega} /V\) normalized on \(G_0 = 2e^2/h \) equals 
\begin{equation}
\label{admit}
		G_{\omega} = \frac{1 - \lambda^2}{\Lambda^2 - \lambda^2} \frac{1}{1 - i \lambda \tan \frac{\omega}{2 \lambda}}
					+ \frac{\Lambda^2 - 1}{\Lambda^2 - \lambda^2} \frac{1}{1 - i \Lambda \tan \frac{\omega}{2 \Lambda}}\,.
\end{equation}
\begin{figure}[htb]
	\centering
	\includegraphics[width=0.9\linewidth]{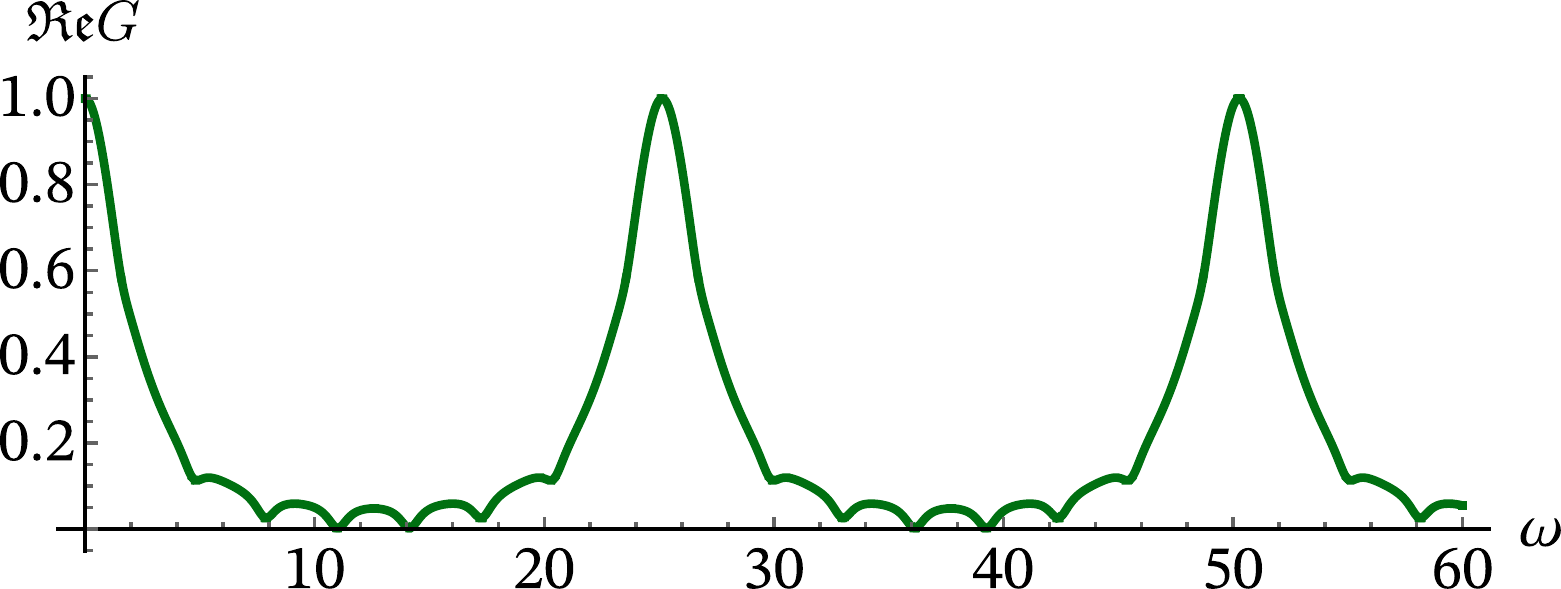}
	\caption{The real part of the admittance versus frequency for electrons with short-range repulsion and iSOI (\(\Lambda = 4\), \(\lambda = 0.5\)).}
	\label{fig1}
\end{figure}
\begin{figure}[htb]
	\centering
	\includegraphics[width=0.9\linewidth]{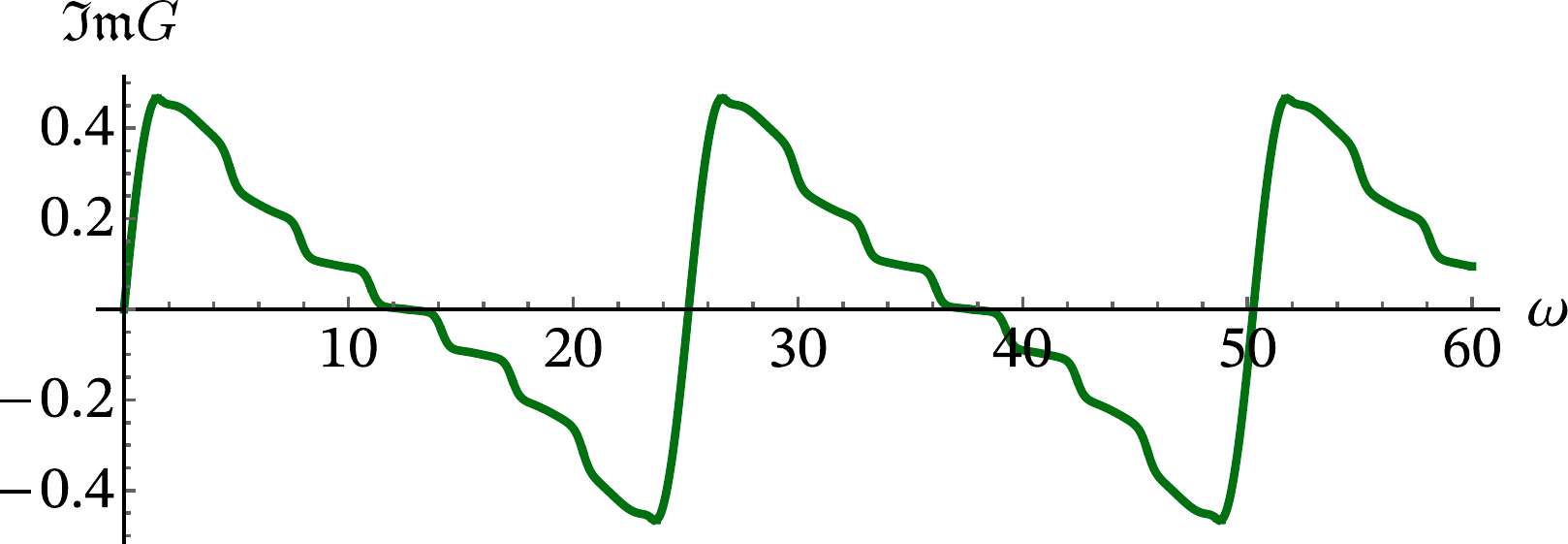}
	\caption{The imaginary part of the admittance versus frequency for electrons with short-range repulsion and iSOI (\(\Lambda = 4\), \(\lambda = 0.5\)).}
	\label{fig2}
\end{figure}

In the limiting case of zero iSOI (\(\lambda = 1\)), this expression reduces to \( G_{\omega} = {\left(1 - i \Lambda \tan \frac{\omega}{2 \Lambda}\right)}^{-1}\), in agreement with Ref.~\cite{PhysRevLett.81.1925}. The real part of the admittance oscillates versus frequency between zero and \(G_0\). At resonant frequencies \(\mathfrak{Re} G\) turns to zero. In this regime electrons in the wire oscillate between the leads, perfectly reflecting from them. Therefore the component of the current that is in phase with the bias voltage vanishes. The resonance condition is \(\omega = \pi (2 n + 1) \Lambda \) for integer \(n\), which means that the frequency is multiple of the inverse transit time of the collective excitation through the quantum wire. The only collective mode contributing to the electron current in the absence of iSOI is the plasmon excitation with the velocity \(\Lambda \), renormalized by the e-e interaction. Consequently, from zeros of \(\mathfrak{Re} G\) one can extract information on how the e-e interaction affects the excitation velocity~\cite{PhysRevLett.81.1925,PhysRevB.58.13847}.

With iSOI present, both modes contribute to the admittance with certain weights. The resulting oscillatory pattern, produced by the interference of the collective modes, has now two different characteristic frequencies, corresponding to different transit times of the slow and fast collective mode, as illustrated by Figs.~\ref{fig1} and~\ref{fig2}. Except for the case of commensurate transit frequencies that can occur provided that \(\Lambda \) and \(\lambda \) are commensurate, \(\mathfrak{Re} G\) no longer turns exactly to zero in its minima. 

The  mode velocities can be determined experimentally from the frequency dependence of the admittance. For example, this can be done using the Fourier analysis of \(G_{\omega}\). The lowest-frequency harmonic is given by
\begin{equation}
	\tilde{G}_{2 \pi \lambda} = \frac{2\lambda}{\Lambda^2 - \lambda^2} \frac{1 - \lambda}{1 + \lambda} \exp \left(i \frac{\omega}{\lambda}\right)\,.
\end{equation}
Notice that in the case of the strong iSOI \(\lambda \ll \Lambda \), which means that the characteristic frequencies in \(G_{\omega}\) are sharply distinct.

\textit{Conclusion}.---The problem of identifying the broken SCS in 1D correlated electron systems was addressed. It was argued that the signatures of the SCS violation should arise in dynamic electron transport. The admittance of a 1D quantum wire coupled to leads was studied to show that its frequency dependence contains the information about the collective modes velocities, affected by the interactions. The spin-charge separation parameter, quantitatively describing the spin-charge structure of the collective excitations, was shown to depend solely on these velocities within the model considered.
 
This work was partially supported by the Russian Foundation for Basic Research (Grant No 17--02--00309) and Russian Academy of Sciences.

\bibliography{paper}

\begin{thebibliography}{20}%
\makeatletter
\providecommand \@ifxundefined [1]{%
 \@ifx{#1\undefined}
}%
\providecommand \@ifnum [1]{%
 \ifnum #1\expandafter \@firstoftwo
 \else \expandafter \@secondoftwo
 \fi
}%
\providecommand \@ifx [1]{%
 \ifx #1\expandafter \@firstoftwo
 \else \expandafter \@secondoftwo
 \fi
}%
\providecommand \natexlab [1]{#1}%
\providecommand \enquote  [1]{``#1''}%
\providecommand \bibnamefont  [1]{#1}%
\providecommand \bibfnamefont [1]{#1}%
\providecommand \citenamefont [1]{#1}%
\providecommand \href@noop [0]{\@secondoftwo}%
\providecommand \href [0]{\begingroup \@sanitize@url \@href}%
\providecommand \@href[1]{\@@startlink{#1}\@@href}%
\providecommand \@@href[1]{\endgroup#1\@@endlink}%
\providecommand \@sanitize@url [0]{\catcode `\\12\catcode `\$12\catcode
  `\&12\catcode `\#12\catcode `\^12\catcode `\_12\catcode `\%12\relax}%
\providecommand \@@startlink[1]{}%
\providecommand \@@endlink[0]{}%
\providecommand \url  [0]{\begingroup\@sanitize@url \@url }%
\providecommand \@url [1]{\endgroup\@href {#1}{\urlprefix }}%
\providecommand \urlprefix  [0]{URL }%
\providecommand \Eprint [0]{\href }%
\providecommand \doibase [0]{http://dx.doi.org/}%
\providecommand \selectlanguage [0]{\@gobble}%
\providecommand \bibinfo  [0]{\@secondoftwo}%
\providecommand \bibfield  [0]{\@secondoftwo}%
\providecommand \translation [1]{[#1]}%
\providecommand \BibitemOpen [0]{}%
\providecommand \bibitemStop [0]{}%
\providecommand \bibitemNoStop [0]{.\EOS\space}%
\providecommand \EOS [0]{\spacefactor3000\relax}%
\providecommand \BibitemShut  [1]{\csname bibitem#1\endcsname}%
\let\auto@bib@innerbib\@empty
\bibitem [{\citenamefont {Manchon}\ \emph {et~al.}(2015)\citenamefont
  {Manchon}, \citenamefont {Koo}, \citenamefont {Nitta}, \citenamefont
  {Frolov},\ and\ \citenamefont {Duine}}]{manchon2015new}%
  \BibitemOpen
  \bibfield  {author} {\bibinfo {author} {\bibfnamefont {A.}~\bibnamefont
  {Manchon}}, \bibinfo {author} {\bibfnamefont {H.~C.}\ \bibnamefont {Koo}},
  \bibinfo {author} {\bibfnamefont {J.}~\bibnamefont {Nitta}}, \bibinfo
  {author} {\bibfnamefont {S.~M.}\ \bibnamefont {Frolov}}, \ and\ \bibinfo
  {author} {\bibfnamefont {R.~A.}\ \bibnamefont {Duine}},\ }\href
  {http://www.nature.com/nmat/journal/v14/n9/full/nmat4360.html} {\bibfield
  {journal} {\bibinfo  {journal} {Nature materials}\ }\textbf {\bibinfo
  {volume} {14}},\ \bibinfo {pages} {871} (\bibinfo {year} {2015})}\BibitemShut
  {NoStop}%
\bibitem [{\citenamefont {Voit}(1995)}]{0034-4885-58-9-002}%
  \BibitemOpen
  \bibfield  {author} {\bibinfo {author} {\bibfnamefont {J.}~\bibnamefont
  {Voit}},\ }\href {http://stacks.iop.org/0034-4885/58/i=9/a=002} {\bibfield
  {journal} {\bibinfo  {journal} {Reports on Progress in Physics}\ }\textbf
  {\bibinfo {volume} {58}},\ \bibinfo {pages} {977} (\bibinfo {year}
  {1995})}\BibitemShut {NoStop}%
\bibitem [{\citenamefont {Moroz}\ \emph {et~al.}(2000)\citenamefont {Moroz},
  \citenamefont {Samokhin},\ and\ \citenamefont {Barnes}}]{PhysRevB.62.16900}%
  \BibitemOpen
  \bibfield  {author} {\bibinfo {author} {\bibfnamefont {A.~V.}\ \bibnamefont
  {Moroz}}, \bibinfo {author} {\bibfnamefont {K.~V.}\ \bibnamefont {Samokhin}},
  \ and\ \bibinfo {author} {\bibfnamefont {C.~H.~W.}\ \bibnamefont {Barnes}},\
  }\href {\doibase 10.1103/PhysRevB.62.16900} {\bibfield  {journal} {\bibinfo
  {journal} {Phys. Rev. B}\ }\textbf {\bibinfo {volume} {62}},\ \bibinfo
  {pages} {16900} (\bibinfo {year} {2000})}\BibitemShut {NoStop}%
\bibitem [{\citenamefont {Gindikin}\ and\ \citenamefont
  {Sablikov}(2017)}]{PhysRevB.95.045138}%
  \BibitemOpen
  \bibfield  {author} {\bibinfo {author} {\bibfnamefont {Y.}~\bibnamefont
  {Gindikin}}\ and\ \bibinfo {author} {\bibfnamefont {V.~A.}\ \bibnamefont
  {Sablikov}},\ }\href {\doibase 10.1103/PhysRevB.95.045138} {\bibfield
  {journal} {\bibinfo  {journal} {Phys. Rev. B}\ }\textbf {\bibinfo {volume}
  {95}},\ \bibinfo {pages} {045138} (\bibinfo {year} {2017})}\BibitemShut
  {NoStop}%
\bibitem [{\citenamefont {Calzona}\ \emph {et~al.}(2015)\citenamefont
  {Calzona}, \citenamefont {Carrega}, \citenamefont {Dolcetto},\ and\
  \citenamefont {Sassetti}}]{PhysRevB.92.195414}%
  \BibitemOpen
  \bibfield  {author} {\bibinfo {author} {\bibfnamefont {A.}~\bibnamefont
  {Calzona}}, \bibinfo {author} {\bibfnamefont {M.}~\bibnamefont {Carrega}},
  \bibinfo {author} {\bibfnamefont {G.}~\bibnamefont {Dolcetto}}, \ and\
  \bibinfo {author} {\bibfnamefont {M.}~\bibnamefont {Sassetti}},\ }\href
  {\doibase 10.1103/PhysRevB.92.195414} {\bibfield  {journal} {\bibinfo
  {journal} {Phys. Rev. B}\ }\textbf {\bibinfo {volume} {92}},\ \bibinfo
  {pages} {195414} (\bibinfo {year} {2015})}\BibitemShut {NoStop}%
\bibitem [{\citenamefont {König}\ \emph {et~al.}(2007)\citenamefont {König},
  \citenamefont {Wiedmann}, \citenamefont {Brüne}, \citenamefont {Roth},
  \citenamefont {Buhmann}, \citenamefont {Molenkamp}, \citenamefont {Qi},\ and\
  \citenamefont {Zhang}}]{Konig766}%
  \BibitemOpen
  \bibfield  {author} {\bibinfo {author} {\bibfnamefont {M.}~\bibnamefont
  {König}}, \bibinfo {author} {\bibfnamefont {S.}~\bibnamefont {Wiedmann}},
  \bibinfo {author} {\bibfnamefont {C.}~\bibnamefont {Brüne}}, \bibinfo
  {author} {\bibfnamefont {A.}~\bibnamefont {Roth}}, \bibinfo {author}
  {\bibfnamefont {H.}~\bibnamefont {Buhmann}}, \bibinfo {author} {\bibfnamefont
  {L.~W.}\ \bibnamefont {Molenkamp}}, \bibinfo {author} {\bibfnamefont {X.-L.}\
  \bibnamefont {Qi}}, \ and\ \bibinfo {author} {\bibfnamefont {S.-C.}\
  \bibnamefont {Zhang}},\ }\href {\doibase 10.1126/science.1148047} {\bibfield
  {journal} {\bibinfo  {journal} {Science}\ }\textbf {\bibinfo {volume}
  {318}},\ \bibinfo {pages} {766} (\bibinfo {year} {2007})}\BibitemShut
  {NoStop}%
\bibitem [{\citenamefont {König}\ \emph {et~al.}(2008)\citenamefont {König},
  \citenamefont {Buhmann}, \citenamefont {Molenkamp}, \citenamefont {Hughes},
  \citenamefont {Liu}, \citenamefont {Qi},\ and\ \citenamefont
  {Zhang}}]{doi:10.1143/JPSJ.77.031007}%
  \BibitemOpen
  \bibfield  {author} {\bibinfo {author} {\bibfnamefont {M.}~\bibnamefont
  {König}}, \bibinfo {author} {\bibfnamefont {H.}~\bibnamefont {Buhmann}},
  \bibinfo {author} {\bibfnamefont {L.~W.}\ \bibnamefont {Molenkamp}}, \bibinfo
  {author} {\bibfnamefont {T.}~\bibnamefont {Hughes}}, \bibinfo {author}
  {\bibfnamefont {C.-X.}\ \bibnamefont {Liu}}, \bibinfo {author} {\bibfnamefont
  {X.-L.}\ \bibnamefont {Qi}}, \ and\ \bibinfo {author} {\bibfnamefont {S.-C.}\
  \bibnamefont {Zhang}},\ }\href {\doibase 10.1143/JPSJ.77.031007} {\bibfield
  {journal} {\bibinfo  {journal} {Journal of the Physical Society of Japan}\
  }\textbf {\bibinfo {volume} {77}},\ \bibinfo {pages} {031007} (\bibinfo
  {year} {2008})}\BibitemShut {NoStop}%
\bibitem [{\citenamefont {Dai}\ \emph {et~al.}(2008)\citenamefont {Dai},
  \citenamefont {Hughes}, \citenamefont {Qi}, \citenamefont {Fang},\ and\
  \citenamefont {Zhang}}]{PhysRevB.77.125319}%
  \BibitemOpen
  \bibfield  {author} {\bibinfo {author} {\bibfnamefont {X.}~\bibnamefont
  {Dai}}, \bibinfo {author} {\bibfnamefont {T.~L.}\ \bibnamefont {Hughes}},
  \bibinfo {author} {\bibfnamefont {X.-L.}\ \bibnamefont {Qi}}, \bibinfo
  {author} {\bibfnamefont {Z.}~\bibnamefont {Fang}}, \ and\ \bibinfo {author}
  {\bibfnamefont {S.-C.}\ \bibnamefont {Zhang}},\ }\href {\doibase
  10.1103/PhysRevB.77.125319} {\bibfield  {journal} {\bibinfo  {journal} {Phys.
  Rev. B}\ }\textbf {\bibinfo {volume} {77}},\ \bibinfo {pages} {125319}
  (\bibinfo {year} {2008})}\BibitemShut {NoStop}%
\bibitem [{\citenamefont {Rothe}\ \emph {et~al.}(2010)\citenamefont {Rothe},
  \citenamefont {Reinthaler}, \citenamefont {Liu}, \citenamefont {Molenkamp},
  \citenamefont {Zhang},\ and\ \citenamefont
  {Hankiewicz}}]{1367-2630-12-6-065012}%
  \BibitemOpen
  \bibfield  {author} {\bibinfo {author} {\bibfnamefont {D.~G.}\ \bibnamefont
  {Rothe}}, \bibinfo {author} {\bibfnamefont {R.~W.}\ \bibnamefont
  {Reinthaler}}, \bibinfo {author} {\bibfnamefont {C.-X.}\ \bibnamefont {Liu}},
  \bibinfo {author} {\bibfnamefont {L.~W.}\ \bibnamefont {Molenkamp}}, \bibinfo
  {author} {\bibfnamefont {S.-C.}\ \bibnamefont {Zhang}}, \ and\ \bibinfo
  {author} {\bibfnamefont {E.~M.}\ \bibnamefont {Hankiewicz}},\ }\href
  {http://stacks.iop.org/1367-2630/12/i=6/a=065012} {\bibfield  {journal}
  {\bibinfo  {journal} {New Journal of Physics}\ }\textbf {\bibinfo {volume}
  {12}},\ \bibinfo {pages} {065012} (\bibinfo {year} {2010})}\BibitemShut
  {NoStop}%
\bibitem [{\citenamefont {Schmidt}\ \emph {et~al.}(2012)\citenamefont
  {Schmidt}, \citenamefont {Rachel}, \citenamefont {von Oppen},\ and\
  \citenamefont {Glazman}}]{PhysRevLett.108.156402}%
  \BibitemOpen
  \bibfield  {author} {\bibinfo {author} {\bibfnamefont {T.~L.}\ \bibnamefont
  {Schmidt}}, \bibinfo {author} {\bibfnamefont {S.}~\bibnamefont {Rachel}},
  \bibinfo {author} {\bibfnamefont {F.}~\bibnamefont {von Oppen}}, \ and\
  \bibinfo {author} {\bibfnamefont {L.~I.}\ \bibnamefont {Glazman}},\ }\href
  {\doibase 10.1103/PhysRevLett.108.156402} {\bibfield  {journal} {\bibinfo
  {journal} {Phys. Rev. Lett.}\ }\textbf {\bibinfo {volume} {108}},\ \bibinfo
  {pages} {156402} (\bibinfo {year} {2012})}\BibitemShut {NoStop}%
\bibitem [{\citenamefont {Rod}\ \emph {et~al.}(2015)\citenamefont {Rod},
  \citenamefont {Schmidt},\ and\ \citenamefont {Rachel}}]{PhysRevB.91.245112}%
  \BibitemOpen
  \bibfield  {author} {\bibinfo {author} {\bibfnamefont {A.}~\bibnamefont
  {Rod}}, \bibinfo {author} {\bibfnamefont {T.~L.}\ \bibnamefont {Schmidt}}, \
  and\ \bibinfo {author} {\bibfnamefont {S.}~\bibnamefont {Rachel}},\ }\href
  {\doibase 10.1103/PhysRevB.91.245112} {\bibfield  {journal} {\bibinfo
  {journal} {Phys. Rev. B}\ }\textbf {\bibinfo {volume} {91}},\ \bibinfo
  {pages} {245112} (\bibinfo {year} {2015})}\BibitemShut {NoStop}%
\bibitem [{\citenamefont {Bocquillon}\ \emph {et~al.}(2013)\citenamefont
  {Bocquillon}, \citenamefont {Freulon}, \citenamefont {Berroir}, \citenamefont
  {Degiovanni}, \citenamefont {Plaçais}, \citenamefont {Cavanna},
  \citenamefont {Jin},\ and\ \citenamefont {Fève}}]{Bocquillon2013}%
  \BibitemOpen
  \bibfield  {author} {\bibinfo {author} {\bibfnamefont {E.}~\bibnamefont
  {Bocquillon}}, \bibinfo {author} {\bibfnamefont {V.}~\bibnamefont {Freulon}},
  \bibinfo {author} {\bibfnamefont {J.-M.}\ \bibnamefont {Berroir}}, \bibinfo
  {author} {\bibfnamefont {P.}~\bibnamefont {Degiovanni}}, \bibinfo {author}
  {\bibfnamefont {B.}~\bibnamefont {Plaçais}}, \bibinfo {author}
  {\bibfnamefont {A.}~\bibnamefont {Cavanna}}, \bibinfo {author} {\bibfnamefont
  {Y.}~\bibnamefont {Jin}}, \ and\ \bibinfo {author} {\bibfnamefont
  {G.}~\bibnamefont {Fève}},\ }\href
  {https://www.nature.com/articles/ncomms2788} {\bibfield  {journal} {\bibinfo
  {journal} {Nature communications}\ }\textbf {\bibinfo {volume} {4}},\
  \bibinfo {pages} {1839} (\bibinfo {year} {2013})}\BibitemShut {NoStop}%
\bibitem [{\citenamefont {Jompol}\ \emph {et~al.}(2009)\citenamefont {Jompol},
  \citenamefont {Ford}, \citenamefont {Griffiths}, \citenamefont {Farrer},
  \citenamefont {Jones}, \citenamefont {Anderson}, \citenamefont {Ritchie},
  \citenamefont {Silk},\ and\ \citenamefont {Schofield}}]{Jompol597}%
  \BibitemOpen
  \bibfield  {author} {\bibinfo {author} {\bibfnamefont {Y.}~\bibnamefont
  {Jompol}}, \bibinfo {author} {\bibfnamefont {C.~J.~B.}\ \bibnamefont {Ford}},
  \bibinfo {author} {\bibfnamefont {J.~P.}\ \bibnamefont {Griffiths}}, \bibinfo
  {author} {\bibfnamefont {I.}~\bibnamefont {Farrer}}, \bibinfo {author}
  {\bibfnamefont {G.~A.~C.}\ \bibnamefont {Jones}}, \bibinfo {author}
  {\bibfnamefont {D.}~\bibnamefont {Anderson}}, \bibinfo {author}
  {\bibfnamefont {D.~A.}\ \bibnamefont {Ritchie}}, \bibinfo {author}
  {\bibfnamefont {T.~W.}\ \bibnamefont {Silk}}, \ and\ \bibinfo {author}
  {\bibfnamefont {A.~J.}\ \bibnamefont {Schofield}},\ }\href {\doibase
  10.1126/science.1171769} {\bibfield  {journal} {\bibinfo  {journal}
  {Science}\ }\textbf {\bibinfo {volume} {325}},\ \bibinfo {pages} {597}
  (\bibinfo {year} {2009})}\BibitemShut {NoStop}%
\bibitem [{\citenamefont {Ulbricht}\ and\ \citenamefont
  {Schmitteckert}(2009)}]{0295-5075-86-5-57006}%
  \BibitemOpen
  \bibfield  {author} {\bibinfo {author} {\bibfnamefont {T.}~\bibnamefont
  {Ulbricht}}\ and\ \bibinfo {author} {\bibfnamefont {P.}~\bibnamefont
  {Schmitteckert}},\ }\href {http://stacks.iop.org/0295-5075/86/i=5/a=57006}
  {\bibfield  {journal} {\bibinfo  {journal} {EPL}\ }\textbf {\bibinfo {volume}
  {86}},\ \bibinfo {pages} {57006} (\bibinfo {year} {2009})}\BibitemShut
  {NoStop}%
\bibitem [{\citenamefont {Maslov}\ and\ \citenamefont
  {Stone}(1995)}]{PhysRevB.52.R5539}%
  \BibitemOpen
  \bibfield  {author} {\bibinfo {author} {\bibfnamefont {D.~L.}\ \bibnamefont
  {Maslov}}\ and\ \bibinfo {author} {\bibfnamefont {M.}~\bibnamefont {Stone}},\
  }\href {\doibase 10.1103/PhysRevB.52.R5539} {\bibfield  {journal} {\bibinfo
  {journal} {Phys. Rev. B}\ }\textbf {\bibinfo {volume} {52}},\ \bibinfo
  {pages} {R5539} (\bibinfo {year} {1995})}\BibitemShut {NoStop}%
\bibitem [{\citenamefont {Safi}\ and\ \citenamefont
  {Schulz}(1995)}]{PhysRevB.52.R17040}%
  \BibitemOpen
  \bibfield  {author} {\bibinfo {author} {\bibfnamefont {I.}~\bibnamefont
  {Safi}}\ and\ \bibinfo {author} {\bibfnamefont {H.~J.}\ \bibnamefont
  {Schulz}},\ }\href {\doibase 10.1103/PhysRevB.52.R17040} {\bibfield
  {journal} {\bibinfo  {journal} {Phys. Rev. B}\ }\textbf {\bibinfo {volume}
  {52}},\ \bibinfo {pages} {R17040} (\bibinfo {year} {1995})}\BibitemShut
  {NoStop}%
\bibitem [{\citenamefont {Sablikov}\ and\ \citenamefont
  {Shchamkhalova}(1998)}]{PhysRevB.58.13847}%
  \BibitemOpen
  \bibfield  {author} {\bibinfo {author} {\bibfnamefont {V.~A.}\ \bibnamefont
  {Sablikov}}\ and\ \bibinfo {author} {\bibfnamefont {B.~S.}\ \bibnamefont
  {Shchamkhalova}},\ }\href {\doibase 10.1103/PhysRevB.58.13847} {\bibfield
  {journal} {\bibinfo  {journal} {Phys. Rev. B}\ }\textbf {\bibinfo {volume}
  {58}},\ \bibinfo {pages} {13847} (\bibinfo {year} {1998})}\BibitemShut
  {NoStop}%
\bibitem [{\citenamefont {Haldane}(1981)}]{haldane1981luttinger}%
  \BibitemOpen
  \bibfield  {author} {\bibinfo {author} {\bibfnamefont {F.~D.~M.}\
  \bibnamefont {Haldane}},\ }\href
  {http://stacks.iop.org/0022-3719/14/i=19/a=010} {\bibfield  {journal}
  {\bibinfo  {journal} {Journal of Physics C: Solid State Physics}\ }\textbf
  {\bibinfo {volume} {14}},\ \bibinfo {pages} {2585} (\bibinfo {year}
  {1981})}\BibitemShut {NoStop}%
\bibitem [{\citenamefont {Shockley}(1938)}]{doi:10.1063/1.1710367}%
  \BibitemOpen
  \bibfield  {author} {\bibinfo {author} {\bibfnamefont {W.}~\bibnamefont
  {Shockley}},\ }\href {\doibase 10.1063/1.1710367} {\bibfield  {journal}
  {\bibinfo  {journal} {Journal of Applied Physics}\ }\textbf {\bibinfo
  {volume} {9}},\ \bibinfo {pages} {635} (\bibinfo {year} {1938})}\BibitemShut
  {NoStop}%
\bibitem [{\citenamefont {Blanter}\ \emph {et~al.}(1998)\citenamefont
  {Blanter}, \citenamefont {Hekking},\ and\ \citenamefont
  {Büttiker}}]{PhysRevLett.81.1925}%
  \BibitemOpen
  \bibfield  {author} {\bibinfo {author} {\bibfnamefont {Y.~M.}\ \bibnamefont
  {Blanter}}, \bibinfo {author} {\bibfnamefont {F.~W.~J.}\ \bibnamefont
  {Hekking}}, \ and\ \bibinfo {author} {\bibfnamefont {M.}~\bibnamefont
  {Büttiker}},\ }\href {\doibase 10.1103/PhysRevLett.81.1925} {\bibfield
  {journal} {\bibinfo  {journal} {Phys. Rev. Lett.}\ }\textbf {\bibinfo
  {volume} {81}},\ \bibinfo {pages} {1925} (\bibinfo {year}
  {1998})}\BibitemShut {NoStop}%
\end{thebibliography}%

\end{document}